\documentclass[useAMS,usenatbib,usegraphicx]{mn2e}

\newcommand{\logm}{\mbox{$\log M_{\star}$}}
\newcommand{\Mstar}{\mbox{$M_{\star}$}}

\newcommand{\ssfr}{\mbox{$SFR/M_{\star}$}}
\newcommand{\gfrac}{\mbox{$M_{\hi}/M_{\star}$}}
\newcommand{\sfe}{\mbox{$SFR/M_{\hi}$}}

\newcommand{\msun}{\mbox{$M_{\odot}$}}

\newcommand{\hi}{H\,{\sevensize I}}
\newcommand{\hicap}{H\,{\sc i}}

\def\deg      {{\ifmmode^\circ\else$^\circ$\fi}} 

\defcitealias{Catinella2010}{Paper I}

\title[GASS II:  The Star Formation Efficiency of Massive Galaxies]{The GALEX Arecibo SDSS Survey II: The Star Formation Efficiency of Massive Galaxies}
\author[D. Schiminovich et al.]
{David Schiminovich$^{1}$\thanks{ds@astro.columbia.edu},
  Barbara Catinella$^{2}$, Guinevere Kauffmann$^{2}$, Silvia Fabello$^{2}$, 
  \newauthor Jing Wang$^{2,3}$, Cameron Hummels$^{1}$, Jenna Lemonias$^{1}$, Sean M. Moran$^{4}$, Ronin Wu$^{5}$,
  \newauthor Riccardo Giovanelli$^{6}$, Martha P. Haynes$^{6}$, Timothy M. Heckman$^{4}$, Antara R. 
  \newauthor Basu-Zych$^{7}$, Michael R. Blanton$^{5}$, Jarle Brinchmann$^{8,9}$, Tam{\'a}s Budav{\'a}ri$^{4}$, Thiago 
  \newauthor Gon{\c c}alves$^{10}$, Benjamin D. Johnson$^{11}$, Robert C. Kennicutt$^{11,12}$, 
  \newauthor Barry F. Madore$^{13}$, Christopher D. Martin$^{10}$, Michael R. Rich$^{14}$, Linda J. Tacconi$^{15}$, 
  \newauthor David A. Thilker$^{4}$, Vivienne Wild$^{16}$, and Ted K. Wyder$^{10}$\\
$^{1}$Department of Astronomy, Columbia University, New York, NY 10027, USA\\
$^{2}$Max-Planck Institut f\"{u}r Astrophysik, D-85741 Garching, Germany\\
$^{3}$Center for Astrophysics, University of Science and Technology of China, 230026 Hefei, China\\
$^{4}$Department of Physics and Astronomy, The Johns Hopkins University, Baltimore, MD 21218, USA\\
$^{5}$Department of Physics, New York University, New York, NY 10003 USA\\
$^{6}$Center for Radiophysics and Space Research, Cornell University, Ithaca, NY 14853, USA\\
$^{7}$NASA Goddard Space Flight Center, Laboratory for X-ray Astrophysics, Greenbelt, MD 20771, USA\\
$^{8}$Leiden Observatory, Leiden University, 2300 RA, Leiden, The Netherlands\\
$^{9}$Centro de Astrof\'{\i}sica, Universidade do Porto, 4150-762 Porto, Portugal\\
$^{10}$California Institute of Technology, Pasadena, CA 91125, USA\\
$^{11}$Institute of Astronomy, Cambridge CB3 0HA, UK\\
$^{12}$Steward Observatory, University of Arizona, Tucson, AZ 85721, USA\\
$^{13}$Observatories of the Carnegie Institution of Washington, Pasadena, CA 91101, USA\\
$^{14}$Department of Physics and Astronomy, University of California, Los Angeles, CA 90095, USA\\
$^{15}$Max Planck Institut f\"{u}r extraterrestrische Physik, D-85741 Garching, Germany\\
$^{16}$Institut d'Astrophysique de Paris, 75014 Paris, France      
}

\begin{document}

\date{}

\pagerange{\pageref{firstpage}--\pageref{lastpage}} \pubyear{2010}

\maketitle

\label{firstpage}

\begin{abstract}
We use measurements of the \hi~content, stellar mass and star formation rates in $\sim$190 massive galaxies with $M_\star > 10^{10}$ \msun, obtained from the Galex Arecibo SDSS survey described in Paper I \citep{Catinella2010} to explore the global scaling relations associated with the bin-averaged ratio of the star formation rate over the \hi\ mass  (i.e., $\Sigma SFR/\Sigma M_{HI}$), which we call the \hi-based star formation efficiency (SFE).   Unlike the mean specific star formation rate, which decreases with stellar mass and stellar mass surface density, the star formation efficiency remains relatively constant across the sample with a value close to $SFE = 10^{-9.5}$ yr$^{-1}$  (or an equivalent gas consumption timescale of $\sim 3 \times 10^9$ yr).  Specifically, we find little variation in SFE with stellar mass, stellar mass surface density, $NUV-r$ color and concentration ($R_{90}/R_{50}$).  We interpret these results as an indication that external processes or feedback mechanisms that control the gas supply are important for regulating star formation in massive galaxies.  An investigation into the detailed distribution of SFEs reveals that approximately 5\% of the sample shows high efficiencies with SFE $> 10^{-9}$ yr$^{-1}$, and we suggest that this is very likely due to a deficiency of cold gas rather than an excess star formation rate.  Conversely, we also find a similar fraction of galaxies that appear to be gas-rich for their given specific star-formation rate, although these galaxies show both a higher than average gas fraction and lower than average specific star formation rate.  Both of these populations are plausible candidates for ``transition'' galaxies, showing potential for a change (either decrease or increase) in their specific star formation rate in the near future.  We also find that $36\pm5$ \% of the total \hi~mass density and $47\pm5$ \% of the total SFR density is found in galaxies with $M_\star > 10^{10}$ \msun.

\end{abstract}


\section{Introduction}

Measurements of volume-averaged stellar mass, star formation rate and gas densities over cosmic time provide fundamental constraints on galaxy evolution by describing the integrated past history, present activity and future potential for galaxy growth. In the past decade, substantial progress has been made in accurately measuring the first two---the stellar mass function \citep{Bell2003,Borch2006} and star formation rate density \citep{Brinchmann2004,Wyder2007,Salim2007}.  On the other hand, quantifying gas content across redshift, and in particular identifying the gas supply and reservoir most closely linked to recent and future star formation---has remained challenging.

Recent progress has been made in the local universe through blind \hi~surveys such as HIPASS \citep{Meyer2004} and ALFALFA \citep{Giovanelli2005} which are yielding \hi~mass functions of representative volumes \citep{Zwaan2005,Stierwalt2009} and optical identifications of nearly all of \hi~detected sources \citep[e.g.,][]{Saintonge2008,Martin2009}.  This latter result, the fact that blind \hi~surveys do not reveal a population of \hi-rich galaxies without optical counterparts, suggests that the study of the gas content of optically selected samples can provide a nearly complete census of \hi~in the local universe.

In this paper, rather than perform a complete census of \hi, we focus exclusively on a subsample of massive galaxies in the local universe that have been homogeneously observed by SDSS, GALEX and now Arecibo as part of the GASS survey \citep[described in ][ hereafter Paper I]{Catinella2010}.  Briefly stated, the goals of the GASS survey are to measure the \hi~properties of $\sim 1000$ massive ($M_\star > 10^{10}$ \msun) galaxies  in the local universe (z$<0.05$), in a mass range that includes the transition mass above which most galaxies have ceased forming stars.  To maximize survey efficiency, galaxies are observed to a gas fraction limit of 2-5\%, ensuring that the quantity of gas probed remains relevant for on-going and future star formation.  In Paper I we found that the \hi~gas fraction decreased with stellar mass, stellar mass surface density and $NUV-r$ color.  

Observations and models suggest that star formation in massive galaxies may be quenched by internal (AGN or SF feedback) or external means (stripping or other environmental effects).  This has been used to explain both the decline of the global star formation rate (SFR) and the growth of stellar mass on the red sequence. If quenching is due to the depletion of a gaseous reservoir, one might expect its signature to be evident in a decrease in the mean \hi~ gas fraction.  However if the quenching is due to internal processes that inhibit star formation \citep[e.g.,][]{Martig2009}, then the signature of quenching might also be reflected in a decreased star formation efficiency (e.g. SFE $= SFR/M_{HI}$) or the equivalent converse, an increased gas consumption timescale ($t_{cons}=M_{HI}/SFR$).  While SFR efficiencies have been discussed extensively \citep[e.g.][]{Roberts1963, Larson1980, Kennicutt1983, Kennicutt1994, Boselli2001, Bothwell2009}, this work marks the first time that it has been determined for a representative sample of galaxies selected exclusively by stellar mass.

In this paper we measure the distribution of SFR efficiencies across the GASS sample and ask whether we find an excess of highly efficient or inefficient star-forming galaxies? We also investigate how the mean efficiency varies across the sample.  Additionally, we take advantage of our simple selection criterion to produce a determination of several volume-averaged physical quantities.  We combine GASS and GALEX measurements of $\sim$190 galaxies with recent determinations of the local stellar mass function to calculate the \hi~mass density and SFR density as a function of stellar mass.  This analysis allows us to compare the properties of our mass-selected sample to that of the full population. We emphasize here that the sample, while restricted to relatively massive galaxies, includes many galaxies that are blue and/or not quiescent, in contrast with samples selected by color or early-type morphology, the latter of which are known to have low gas fractions and low star formation rates \citep[e.g.,][]{Bregman1992,  Yi2005, Morganti2006}. In fact, as we discuss below, a significant fraction of the total star formation rate in the local universe is taking place within galaxies in this stellar mass range.  

We use these measurements to address several fundamental questions: How much of the total \hi\ in the local universe is associated with massive galaxies?  How does this change across the so-called transition mass of $10^{10.5}$ \msun?  How does the \hi\ density compare with the measured SFR density over the same mass range, and what does this suggest regarding the global SFR efficiency of massive galaxies? What fraction of this gas is in the process of efficiently forming stars, vs. the fraction that may be building up a reservoir for future star formation?      We interpret our results in the context of quenching models, as well as scenarios that may lead to the return of a galaxy back onto the star-forming sequence.

Throughout this paper, we make use of the flat $\Lambda$CDM cosmology with $H_0$ = 70 km s$^{-1}$ Mpc$^{-1}$ and $\Omega_\Lambda =0.7$.

\section{Data}

\subsection{\hi, Optical Data and Derived Quantities}

Our \hi~data are taken from the GASS first data release described in Paper I and we refer the reader there for details on the observations and initial data analysis.   We briefly review the most pertinent information. Our large parent sample (PS) contains 12006 galaxies with $M_\star > 10^{10}$ \msun~and 0.025$<$z$<$0.05 visible from Arecibo and located within the footprint of the SDSS primary spectroscopic survey,  the projected GALEX Medium Imaging Survey (MIS) and ALFALFA.   The first release of GASS data contains $\sim190$ galaxies of which 176 are new measurements, with 99 detections and 77 upper limits. Another $\sim10-15$ previously detected gas-rich galaxies (from ALFALFA and other surveys) are added to produce the statistically representative, volume-limited sample (DR1) used in this paper.   DR1 contains  20\% of the galaxies planned for the full GASS survey.   

\hi~masses are calculated for the detected galaxies, and upper limits determined for the non-detections.  For this measurement our targets are considered to be unresolved by the Arecibo 3.5$^\prime$ beam.  Additionally, no correction is made for self-absorption. Detected \hi~masses range from 4.6$\times 10^8$ to 3.2$\times 10^{10} M_\odot$.   Upper limits (5$\sigma$) from non-detections assume a 300 km/s velocity width.  Upper limits are indicated on figures using arrows plotted at the location of the 5$\sigma$ detection limit.  Although velocity widths have been measured from the Arecibo spectrum, we do not make use of those quantities in this paper.    We have investigated possible source confusion and contamination due to signal from galaxies close to or at the same redshift as our target galaxies.  While such confusion does not have an influence on our main results, where appropriate we have indicated those galaxies for which confusion might cause an overestimate of the object's \hi~mass.
 
Although the GASS parent sample was defined using SDSS DR6, measurements and derived physical properties reported here were obtained using SDSS DR7, including the MPA/JHU value-added catalogs. Ultraviolet photometric measurements were calculated directly from pipeline-processed GALEX images based on GALEX data release GR6.   These quantities have been tabulated in Paper I.   All photometric quantities used in this paper have been corrected for Galactic extinction. Stellar masses have been calculated using SDSS photometry only using the methodology described in \citet{Salim2007}.   Following that work, we assume a \citet{Chabrier2003} IMF for all derived quantities based on stellar masses and star formation rates, including gas mass fractions and star formation rate efficiencies.    Stellar mass surface densities are calculated assuming that 50\% of the stellar mass is contained within the r-band half-light radius.

\begin{figure*}
\includegraphics[width=168mm]{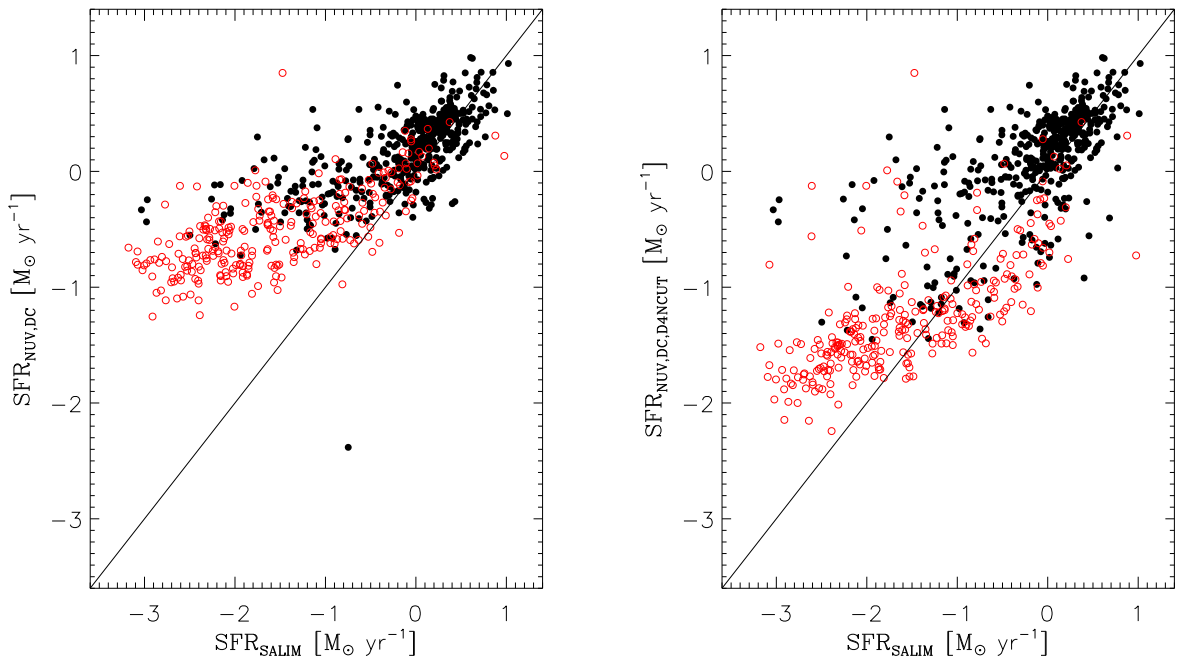}
\caption{Comparison of SFR derived using methodology in this paper and SFR from \citet{Salim2007}, illustrating how SFRs bracket a range of possible values. In both plots, black filled points are galaxies with SDSS D$_n$(4000) $<$ 1.7 and red open circles with D$_n$(4000) $>$ 1.7.  Left panel shows comparision of dust-corrected SFR for full range of galaxies (DC sample), based on \citet{Johnson2007}.  Right panel shows comparison with dust-corrected SFR with a D$_n$(4000) cut on attenuation (DC,D4NCUT sample) , meaning that A$_{NUV}$ = 0 for galaxies with D$_n$(4000) $>$ 1.7.  }
\label{Fig:sfr_compare}
\end{figure*}

\begin{figure*}
\includegraphics[width=135mm]{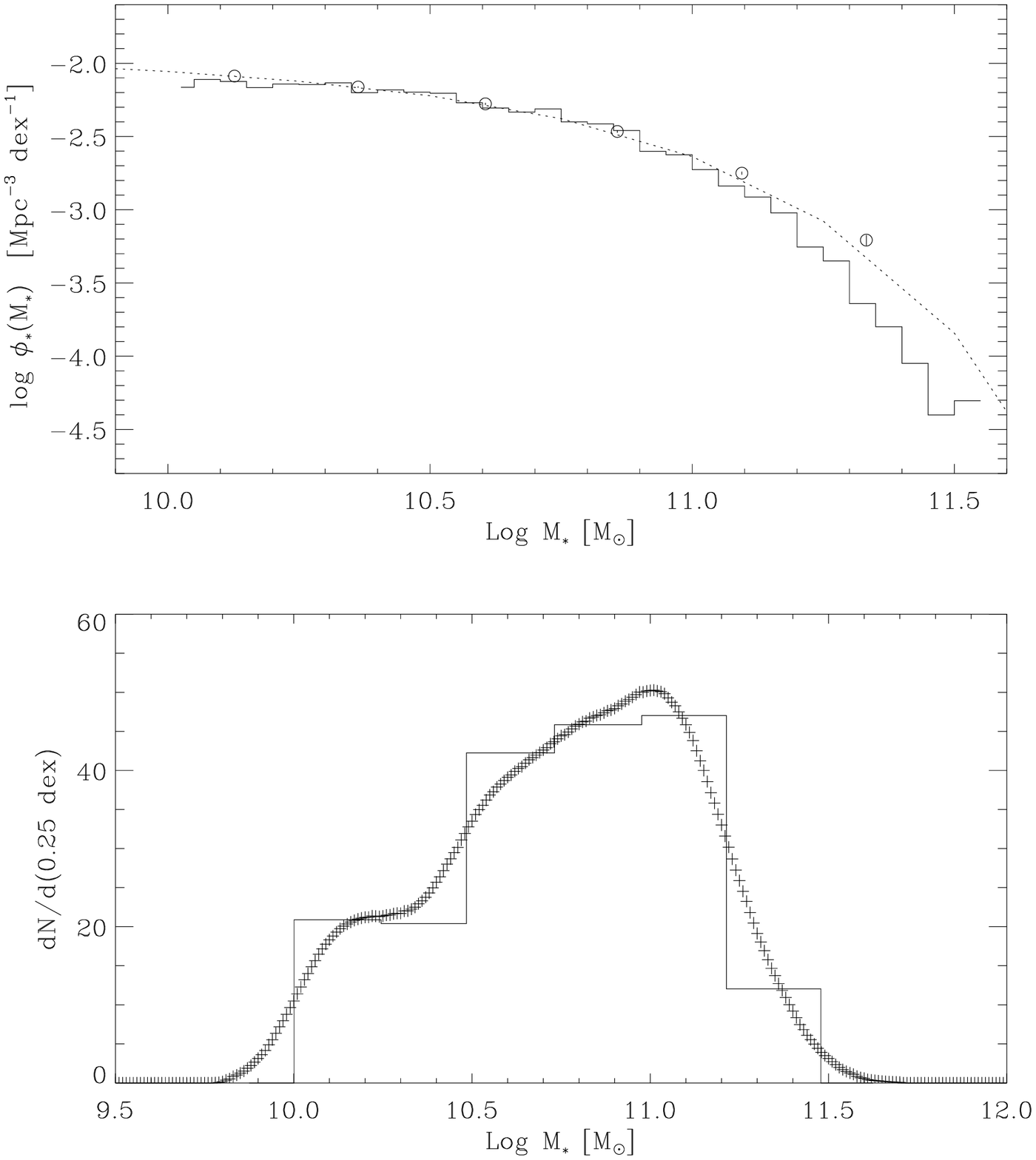}
\caption{Stellar mass function and distribution of GASS sample galaxies.  (Top panel)  Circles show values calculated in 0.25 dex bins using the methodology described in the paper with bootstrap errors.  Solid histogram shows distribution of GASS PS galaxies selected from SDSS DR6, with normalization scaled to fit stellar mass function. Dotted curve is stellar mass function from \citet{Borch2006}.  (Bottom panel) Histogram shows binned stellar mass distribution from GASS DR1 sample.  Plus signs show Gaussian-smoothed density function vs. M$_\star$ for DR1.}
\label{Fig:MF}
\end{figure*}

\subsection{Star Formation Rates and Star Formation Efficiencies}

In this section we describe the method we use to calculate star formation rates and star formation efficiencies.  Star formation rates in this paper are derived from UV luminosities corrected for internal dust-attenuation.   A particular challenge is that the GASS sample contains many galaxies with low level star-formation activity for which old and/or non-star-forming components (evolved stars, AGN) can produce a UV luminosity comparable to those of a faint young stellar population.  This issue is not unique to the ultraviolet---deriving SFRs for nearly passive galaxies is challenging using any star formation diagnostic (e.g. UV, IR, H$\alpha$, radio continuum, etc.).  Furthermore, measuring dust attenuation in individual galaxies is a complex problem, and nearly unfeasible with low S/N data.    A full treatment of these issues is beyond the scope of this paper, and we adopt a relatively simple UV-optical based approach following analyses in previous work.  Interesting alternative multi-wavelength methods for deriving SFRs across the galaxy population have been developed \citep{Salim2007, Kennicutt2009, Cortese2008} and we will explore these in future work.

A simple, one-component expression for calculating star formation rates using observed UV luminosities is given by:
$$SFR = \frac{L_{UV} f_{UV}(young) 10^{0.4A_{UV}}}{\eta_{UV}},$$
where $\eta_{uv}$ is the (star-formation history and metallicity-dependent) conversion factor between UV luminosity and recent-past-averaged star formation rate,  A$_{UV}$ is the (geometrically-averaged) ratio of intrinsic UV luminosity to measured UV luminosity, and f$_{UV}$(young) is a measure of the fraction of UV light that originates in a young---as opposed to a highly evolved stellar population.  All of these quantities are highly simplified and cannot be considered to be independent, but for our derivations and discussion below they will be treated as such.

Since the GASS sample is drawn from SDSS and therefore has 7 band photometry ($fnugriz$) and spectral line indices, we use dust attenuations (A$_{NUV}$) derived using the methodology described in \citet{Johnson2007}, slightly modified to produce more accurate SFRs for galaxies with evolved stellar populations. Variations on this methodology have been proposed \citep[e.g.][]{Cortese2008} and we have checked this method against ours to ensure sure that applying different corrections has little impact on our conclusions.  We do not  employ the ``hybrid'' attenuation correction used in \citet{Schiminovich2007} because of the unavailability of the z-band dust attenuation measure ($A_z$), although we have calculated similar dust attenuation measures using $\tau_V$ in order to verify the robustness of the results reported here.  We also do not make use of H-alpha derived star formation rates because emission lines with sufficient S/N are only available for a subset of the sample and fiber aperture corrections can be large for our (relatively) low-redshift sample.

\citet{Johnson2006, Johnson2007} used the IR/UV flux ratio (IRX) from a sample of 1000 SDSS galaxies to derive a UV-based dust attenuation measure ($A_{IRX}$). By combining UV-optical colors with D$_n$(4000), which correlates with star formation history, \citet{Johnson2007} showed that:
$$A_{IRX} = 1.25 -1.33x+1.19 y-1.02 x y$$
where $x= D_n(4000)-1.25$ and $y=~^{0.1}(NUV-r)-2$ and coefficients have been taken from Table 2 in \citet{Johnson2007}.    \citet{Johnson2007} show that the empirically derived $A_{IRX}$ shows a close correspondence to $A_{UV}$ for galaxies with D$_n$(4000)$<$1.7 and using the  \citet{Calzetti2000} attenuation curve one can show that $A_{NUV}= 0.81 A_{IRX}$. 

For galaxies with D$_n$(4000)$>$1.7 the IR/UV flux ratio is also sensitive to dust heating by light from evolved stars and is not a good measure of the dust attenuation of a young stellar population.  We attempt to bound the range of possible dust attenuations by considering two distinct cases.  In the first (DC,D4NCUT), we assume that the IR luminosity in galaxies with D$_n$(4000)$>$1.7 is reprocessed light from an old population and we therefore set A$_{NUV}=0$ for galaxies with D$_n$(4000)$>$1.7.  In the second case (DC), we apply our empirically derived dust attenuation corrections to the entire sample.  

We make a further simplification by assuming that f$_{UV}$(young) = 1, or that all of the UV light in our measurements comes from a young stellar population.  Measurements of the UV-optical colors of early-type galaxies \citep[e.g.,][]{Rich2005,Yi2005,Donas2007} show NUV-r $\sim$5-6.  Such a color, converted into a specific star formation rate, leads to a value close to \ssfr $\sim 10^{-12}$ yr$^{-1}$.  This implies that specific star formation rates close to this value will very likely include a contribution from old stars and will most likely lead to an overestimate of SFR, since f$_{UV}$(young) may be much less than 1.  While our assumption leads to an overestimate of SFRs  in weakly star-forming galaxies, it ends up having a small effect on average quantities discussed in the first part of this paper.  

Star formation rates were calculated assuming a constant $\eta_{NUV} = 10^{28.165}$ and therefore: 
$$ SFR {\rm (M_\odot/yr)} =  10^{-28.165} L_{\nu,NUV} {\rm (erg ~ s^{-1} Hz^{-1}) 10^{0.4A_{NUV}}}$$ as derived by \citet{Salim2007} assuming a Chabrier (2003) IMF and a continuous recent (100-300 Myr) star formation history, which makes these star formation rates directly comparable to those in that work and most other recent determinations.  For a standard Salpeter IMF (between 0.1 and 100 $M_\odot$), star formation rates would be a factor of $\sim1.5$ higher.     

In this paper we are primarily focused on quantities derived using the ratio of mean values.  In several cases, but primarily for the star formation rates, these mean quantities can be dominated by a few very high star formation rate galaxies.  Therefore, we took some care to ensure that our results were not affected by possible errors in measured SFRs for galaxies with the highest SFR.  Many of these galaxies are IR-luminous and detected by IRAS.  We have compared our derived star-formation rate and the total infrared luminosity, obtained from the IIFSCz \citep[Imperial IRAS FSC redshift catalog;][]{Wang2009}.  Overall, the agreement is sufficiently good that differences are unlikely to affect our results.  It is possible, however, that we could be underestimating the SFR for the most IR-luminous galaxies, thereby underestimating the star formation efficiencies for a subset of our sample.

In Figure \ref{Fig:sfr_compare} we compare SFRs obtained for both cases described above, to a subset of GASS parent sample galaxies with SFRs calculated previously by \citet{Salim2007}.  We note that the DC,D4NCUT sample compares quite favorably to \citet{Salim2007} and we adopt this as the primary sample for this paper.

Finally,  \hi~ masses, stellar masses and star formation rates are used to derive \hi-based gas fractions ($M_{\hi}/M_\star$), star formation efficiencies (SFE=\sfe) and specific star formation rates (\ssfr).  We follow \citet{Leroy2008} in adopting SFE, the SFR-to-gas ratio, as our primary derived quantity, as opposed to its reciprocal, often called the gas consumption timescale (or the ``Robert's time'').  Although our definition is straightforward, it is not necessarily identical to the star formation efficiency term sometimes used in star formation laws and related prescriptions.

For all of our derived quantities we have opted for the simplest possible definition.  We do not include a gas recycling term when calculating star formation efficiency.  Although helium adds 26\% to the cold gas mass we do not include this factor to calculate a cold gas mass, nor do we include a correction for the unmeasured molecular component.  This is in contrast, e.g. to the factor of 2.3 applied to \hi~masses in the SINGG survey \citep{Meurer2006,Hanish2006}.  We do note that the molecular phase in $\sim$1/3 the GASS sample is currently being studied in a corollary CO survey being carried out at IRAM (COLDGASS, Kauffmann, Kramer, Saintonge 2010).\footnote{http://www.mpa-garching.mpg.de/COLD\_GASS}

\section{Global \hi, SFR and SFE Density}

\subsection{Methodology}

The very simple selection criterion used for the GASS PS allows us to combine our DR1 \hi\ measurements with the local stellar mass function to determine the  volume-averaged \hi~mass density for massive ($M_\star > 10^{10}$ \msun) galaxies in the local universe.  Using SFR measurements we can also determine the analogous SFR density and volume-averaged SFE.  These volume-weighted quantities are derived using the local stellar mass function $\phi_\star(M_\star)$ taken from  \citet{Borch2006}, under the assumption that GASS DR1 observations sample an unbiased, volume-limited distribution of galaxies in any particular stellar mass bin.  

We first derive weights that allow us to scale the GASS sample to match the stellar mass function of \citet{Borch2006}.  We also show that our weighting scheme allows us to (trivially) recover $\phi_\star(M_\star)$ using the GASS DR1 sample.   To avoid effects that might arise from binning, we express the distribution of stellar masses in GASS DR1 as a continuous (Gaussian-smoothed) density function $d(M\star)$, shown in the bottom panel of Figure \ref{Fig:MF}.    

For all $i$ such that $ |\log M_{\star,i} - \log M_{\star,bin}| < \Delta \log M_{\star,bin}/2 $:
$$\phi_\star ( \bar{M}_{\star,bin}) = \frac{\sum_i \phi_\star(M_{\star,i})w_i }{\sum_i w_i} $$ where
$$w_i = \frac{\phi_\star(M_{\star,i})}{d(M_{\star,i})}$$ is a weight that corresponds to the inverse of the effective survey volume and $\Delta \log M_{\star,bin}$ is the bin width (typically 0.25).   
We also define $$\bar{M}_{\star,bin} =  \frac{\sum_i M_{\star,i}w_i}{\sum_i w_i}$$ as the weighted-average stellar mass of any given bin.  

In the top panel of Figure \ref{Fig:MF} we plot the local stellar mass function from \citet{Borch2006} and compare it with $\phi_\star ( \bar{M}_{\star,bin}) $ based on GASS DR1, which we find to be consistent with the input stellar mass function.   We also plot the distribution of stellar masses in the (volume-limited) GASS PS which shows a deficit at high stellar masses, likely due to SDSS spectroscopic targeting criteria that avoid bright objects.  Our method corrects for this incompleteness, which in any case is only significant for the highest stellar masses and is unlikely to impact our analysis.

We can use the same methodology to calculate $\rho_{SFR}(M_\star)$ and $\rho_{HI}(M_\star)$, the volume densities  of \hi~ and SFR as a function of M$_\star$. Again, because we are complete in any given $M_\star$ bin we find that:
$$\rho_{SFR} ( \bar{M}_{\star,bin}) = \frac{\sum_i SFR_i \phi_\star(M_{\star,i})w_i }{\sum_i w_i} $$  and 
$$\rho_{HI} ( \bar{M}_{\star,bin}) = \frac{\sum_i M_{HI,i} \phi_\star(M_{\star,i})w_i }{\sum_i w_i} $$  

Lastly, we apply a similar technique to determine the contribution of GASS galaxies to the overall \hi~mass density function $\rho_{HI} (M_{HI})$ and SFR density function
$\rho_{SFR}(SFR)$ There are two caveats that should be noted with this calculation:

\begin{enumerate}

\item The GASS sample will not necessarily be complete in any given $M_{HI}$ or SFR bin, and therefore the derived function should only be taken as a lower bound on the total function. 

\item If we perform the calculation by setting the \hi\ mass for non-detections to the \hi~upper limit, we may misrepresent the shape of the density function below the point where the GASS sample is nearly complete.  Because this calculation can provide an approximate upper bound on the actual function, we show these for illustrative purposes. However, these upper-limit mass functions should be treated with caution.  

\end{enumerate}

If we consider a complete bivariate distribution function, e.g. $\phi(M_{HI},M_\star)$ we can evaluate a ``partial'' distribution function:
$$ \phi (M_{HI}, >M_{\star,lim}) = \int_{M_{\star,lim} }^\infty \phi(M_{HI},M_\star) dM_\star $$
where we evaluate this function for a given stellar mass limit (e.g. $M_{\star,lim}=10^{10}$ \msun\ for GASS).  This partial distribution function can be simply calculated using our weight function derived above.  Summing over all galaxies $j$ where $ |\log M_{HI,j} - \log M_{HI,bin}| < \Delta \log M_{HI,bin}/2 $ and stellar mass $M_{\star,j} >M_{\star,lim}$, we find that
$$\phi (\bar{M}_{HI,bin},>M_{\star,lim}) = \sum_j w_j. $$ As discussed above, we calculate this in two different ways to bracket our characterization of the function, either by omitting non-detections, or by setting their \hi\ mass to our calculated upper limit.

    The partial \hi\ mass density function is then $$\rho (M_{HI},>M_{\star,lim})= M_{HI} \phi (M_{HI},>M_{\star,lim})$$ and integrating, we obtain the cumulative function
$$\rho_{HI} (>M_{HI,lim},>M_{\star,lim}) = $$
$$=\int_{M_{HI,lim}}^\infty M_{HI}^\prime \phi (M_{HI}^\prime, >M_{\star,lim}) dM_{HI}^\prime.$$
The integral of the cumulative \hi\ mass density function provides an estimate of the total contribution of GASS galaxies to the \hi\ density in the local universe.

A similar calculation applies for $\phi(SFR,M_\star)$.

\begin{figure*}
\includegraphics[width=168mm]{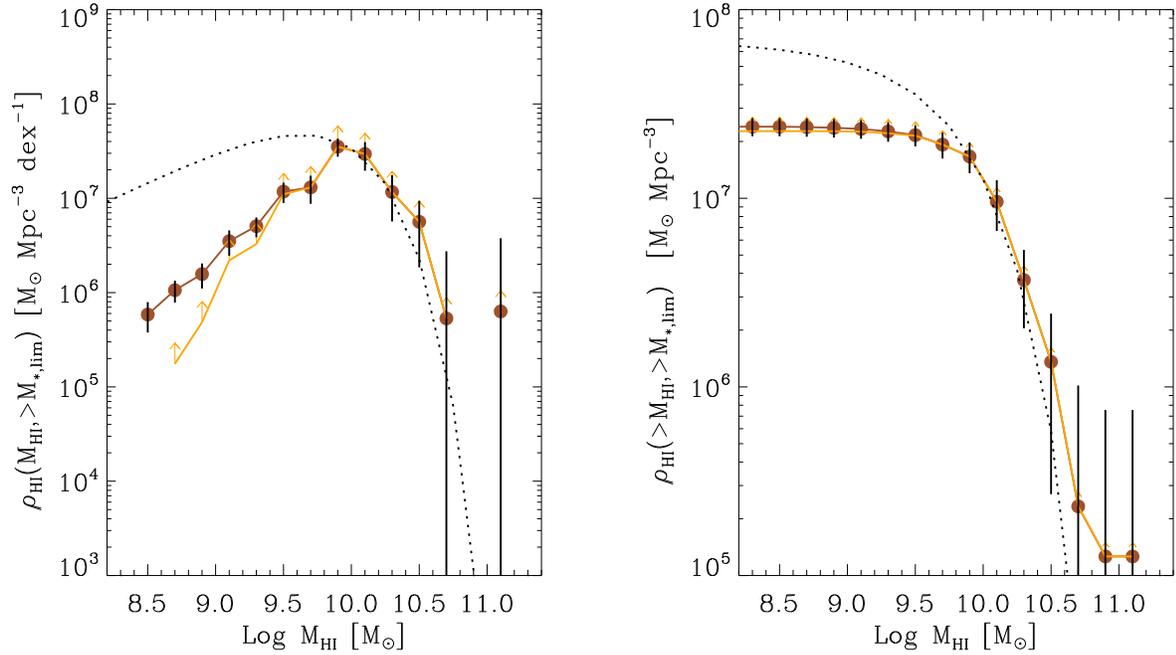}
\caption{Derivation of cumulative \hicap~mass found in massive galaxies (M$_\star > 10^{10}$ \msun).   Yellow curve shows function derived using only \hi~detections and can be considered a lower limit. Brown curve and points show same function where we have included upper limits, which provides some indication of the upper bound. Dotted line shows total \hicap~mass function derived from \citet{Zwaan2005}. (Left) \hicap~mass density vs. \hicap~mass (Right) Cumulative \hicap~mass density above a given \hi~mass. }
\label{Fig:HI}
\end{figure*}

\begin{figure*}
\includegraphics[width=168mm]{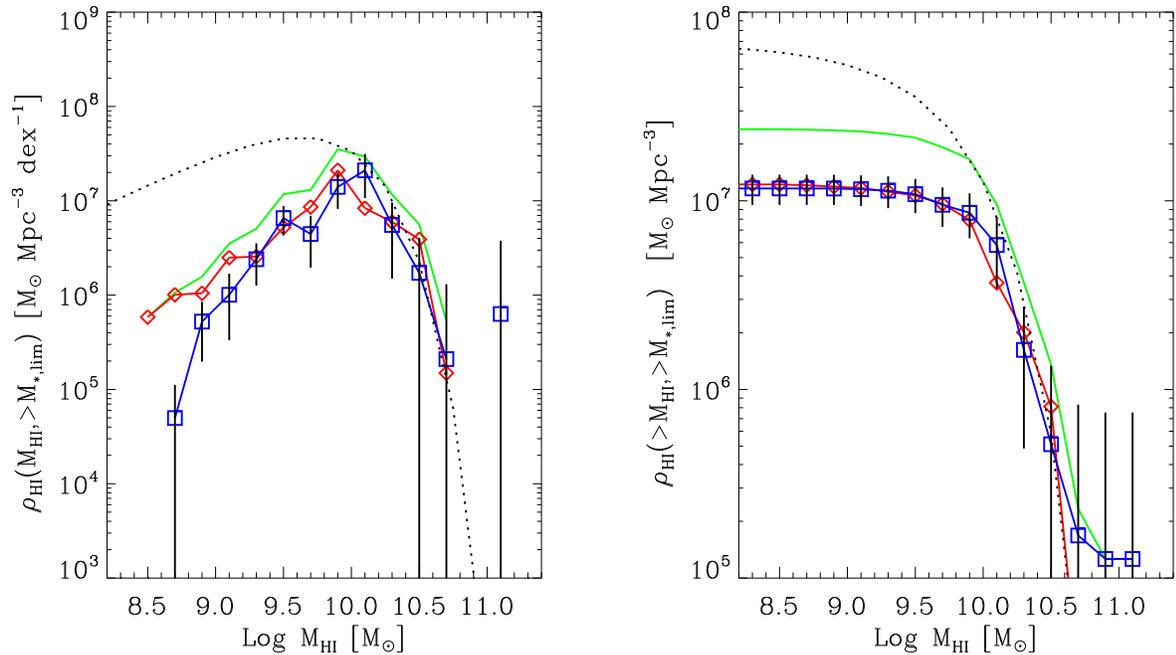}
\caption{Derivation of cumulative \hicap~mass found in massive galaxies (\logm$>$ 10) split by concentration. (Left) \hicap~mass density vs. \hicap~mass (Right) Cumulative \hicap~mass density above a given \hi~mass.   Green histogram indicates full sample, blue histogram low concentration ($R_{90}/R_{50}$ $<$2.6) subsample, red histogram, high concentration ($R_{90}/R_{50}$ $>$2.6)  Dotted line derived from \citet{Zwaan2005}.  Error bars shown only for low concentration subsample.}
 \label{Fig:HIsub}
\end{figure*}

\subsection{Volume-Averaged \hi, SFR, SFE}

We show the partial \hi\ mass density function calculated for the GASS sample in Figure \ref{Fig:HI}, and compare it to the total \hi\ mass density distribution calculated using the \citet{Zwaan2005} \hi\ mass function.  In both plots we show the \hi\ mass density derived using only \hi~detections (yellow curve), which can be taken to be a lower limit, and the \hi\ mass density using upper limits as discussed above (brown curve with points).  Interestingly the GASS-derived curve matches the total \hi~mass density down to $M_{HI}\sim 10^{10}$ \msun, suggesting that nearly all galaxies with high \hi-masses have stellar masses with $M_\star>10^{10}$ \msun.  The GASS sample becomes increasingly less complete for \hi\ masses below this limit.

We also find that GASS galaxies with $ \log M_{HI} > 9.75$  provide the bulk of the contribution to the total \hi~mass density, falling off significantly at lower gas masses.  This can also be seen in the sharp rise in the cumulative \hi\ mass density $\rho_{HI} (>M_{HI},>M_{\star,lim})$ over this \hi\ mass range.
 The cumulative mass density levels off at $(2.5\pm0.3)\times10^7$ M$_\odot$ Mpc$^{-3}$, $36\pm5$\% of the total \hi\ mass density derived using \citet{Zwaan2005}.  This integrated result shows that a significant fraction of the \hi\ mass in the local universe is associated with massive galaxies.  There is little difference between the two curves, indicating that this result is robust with respect to the method that we use to account for non-detections.

We explore in Figure \ref{Fig:HIsub} how this distribution changes if we split our sample according to a galaxy's concentration, the ratio $R_{90}/R_{50}$, which we use as a proxy for distinguishing disk-dominated ($R_{90}/R_{50}$ $<$2.6) and bulge-dominated ( $R_{90}/R_{50}$ $>$2.6) galaxies  \citep[as in e.g.][]{Kauffmann2003}.    This cut yields 54 (133) low (high) concentration galaxies.  We find that this split produces two nearly identical distributions, both contributing nearly equally to the total integrated \hi~mass density.  For the GASS stellar mass range, disk-dominated galaxies do not account for the majority of the gas content of galaxies.  Instead, \hi\  appears to be evenly distributed across a range of galaxy types.

From our derived SFR density function we measure a partial SFR density of ($8.5\pm0.5) \times10^{-3}$ M$_\odot$ Mpc$^{-3}$ yr$^{-1}$ which is 47$\pm$4\% of the total SFR density \citep{Salim2007, Wyder2007} and consistent with the results in those papers measured over the GASS stellar mass range. Interestingly this suggests that GASS galaxies account for nearly the same fraction of the total SFR density as they do in \hi\ content.  We can also calculate an integrated volume-averaged SFE and implied gas consumption timescale which we determine to be $2.9\times10^{-10}$ yr$^{-1}$ (timescale = 3.4$\pm$0.4 Gyr) for GASS galaxies ($M_\star > 10^{10}$ \msun).  Because we have measurements of the total \hi\ and SFR density, we can calculate a SFE for lower mass galaxies ($M_\star < 10^{10}$ \msun), which we find to be $2.1\times10^{-10}$ yr$^{-1}$ (timescale = 4.6$\pm$0.5 Gyr).  The volume-averaged SFE appears to be relatively constant across the full galaxy population.

\begin{figure*}
\includegraphics[width=168mm]{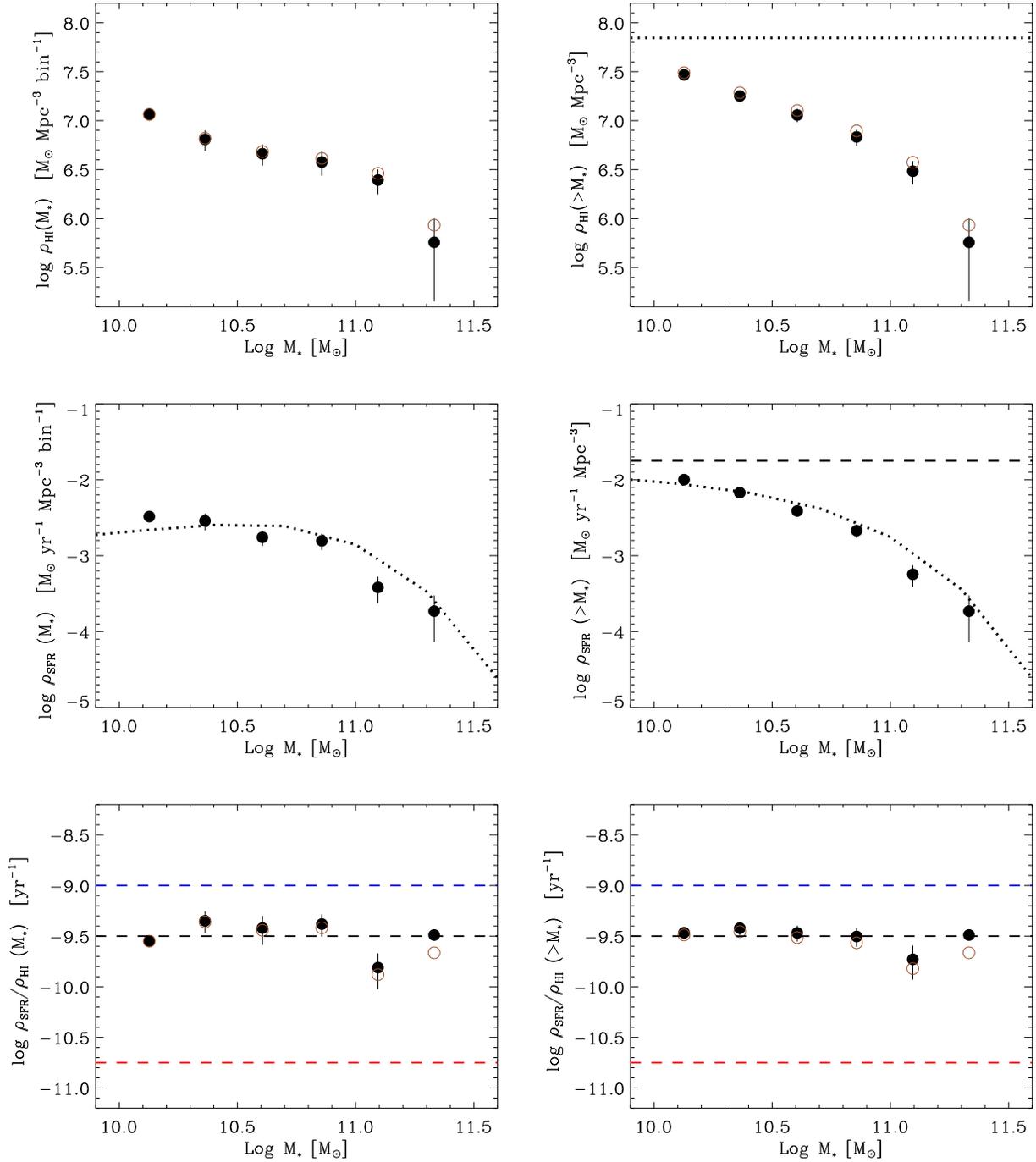}
\caption{\hicap~ and SFR density and volume-averaged SFE vs. \logm.  For upper and lower panels, black dots show results derived from \hi\ detections only, brown dots are derived including non-detections, with \hi\ masses set to calculated upper limits. {\it Left:} Densities shown per \logm~bin (0.25 dex).  {\it Right:} Cumulative density above a given \Mstar.  {\it Top:} \hicap~ density, $\rho_{HI}$.   Dotted line indicates cumulative \hicap~ density derived from \citet{Springob2005} {\it Middle:} SFR density.  Dotted lines show SFR density, $\rho_{SFR}$ vs. \Mstar~from \citet{Schiminovich2007}.  {\it Middle Right:} Dashed line indicates total value from \citet{Salim2007}.  {\it Bottom:} Star formation efficiencies (or inverse gas consumption timescales).  Blue, black and red horizontal dashed lines correspond to high, average and low SFE values found for individual galaxies. }
\label{Fig:SFEdist}
\end{figure*}

\subsection{Global scaling relations: sSFR and SFE}

Although our results suggest that the average SFE is similar above and below $10^{10}$ \msun, it is possible that it only changes significantly above the transition mass at $M_\star \sim 3\times 10^{10}$ \msun.  We show in Figure \ref{Fig:SFEdist} how the volume-averaged \hi~mass, SFR and SFE density functions vary with M$_\star$. Quantities are plotted per M$_\star$ bin on the left, and as a cumulative quantity on the right.  In the bottom panels, we see that the trend in gas consumption remains quite flat across all stellar masses, remaining constant up to our highest stellar mass bin, though with increasing scatter.

\begin{figure*}
\includegraphics[width=168mm]{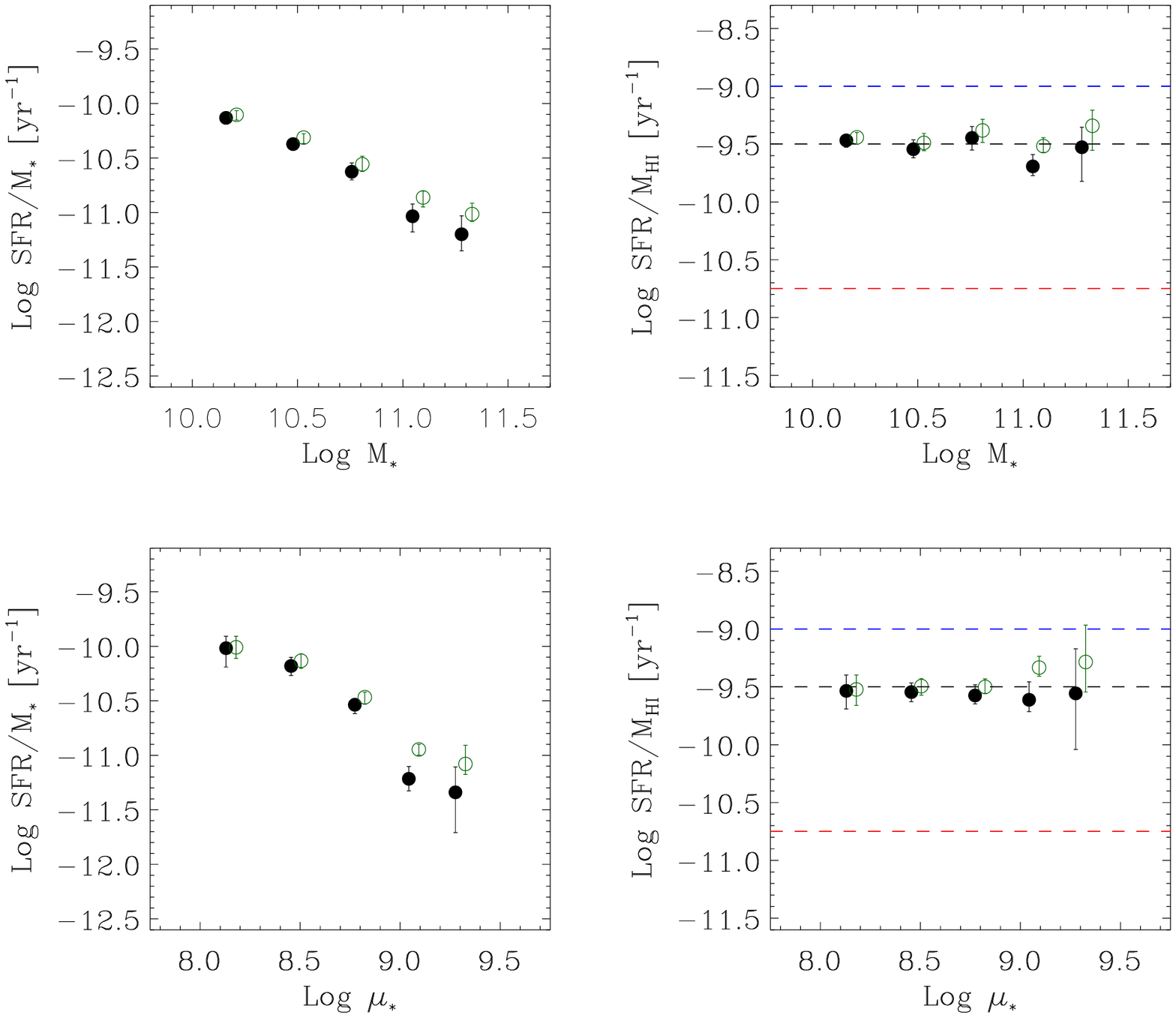}
\caption{Mean specific star formation rate (left) and star formation efficiency (right) as a function of derived quantities stellar mass, M$_\star$ (top) and stellar mass surface density $\mu_\star$ (bottom).  Black points use star formation rates calculated using dust correction $A_{NUV}$ for ``star-forming'' galaxies only (DC,D4NCUT sample), green points apply a dust-correction over the full sample (DC sample).  Error bars (1$\sigma$) derived from bootstrap resampling.  Blue, black and red horizontal dashed lines correspond to high, average and low SFE values found for individual galaxies.   \hi\ non-detections are given zero gas mass.  See text for details.}
\label{Fig:sfe_all}
\end{figure*}

\begin{figure*}
\includegraphics[width=168mm]{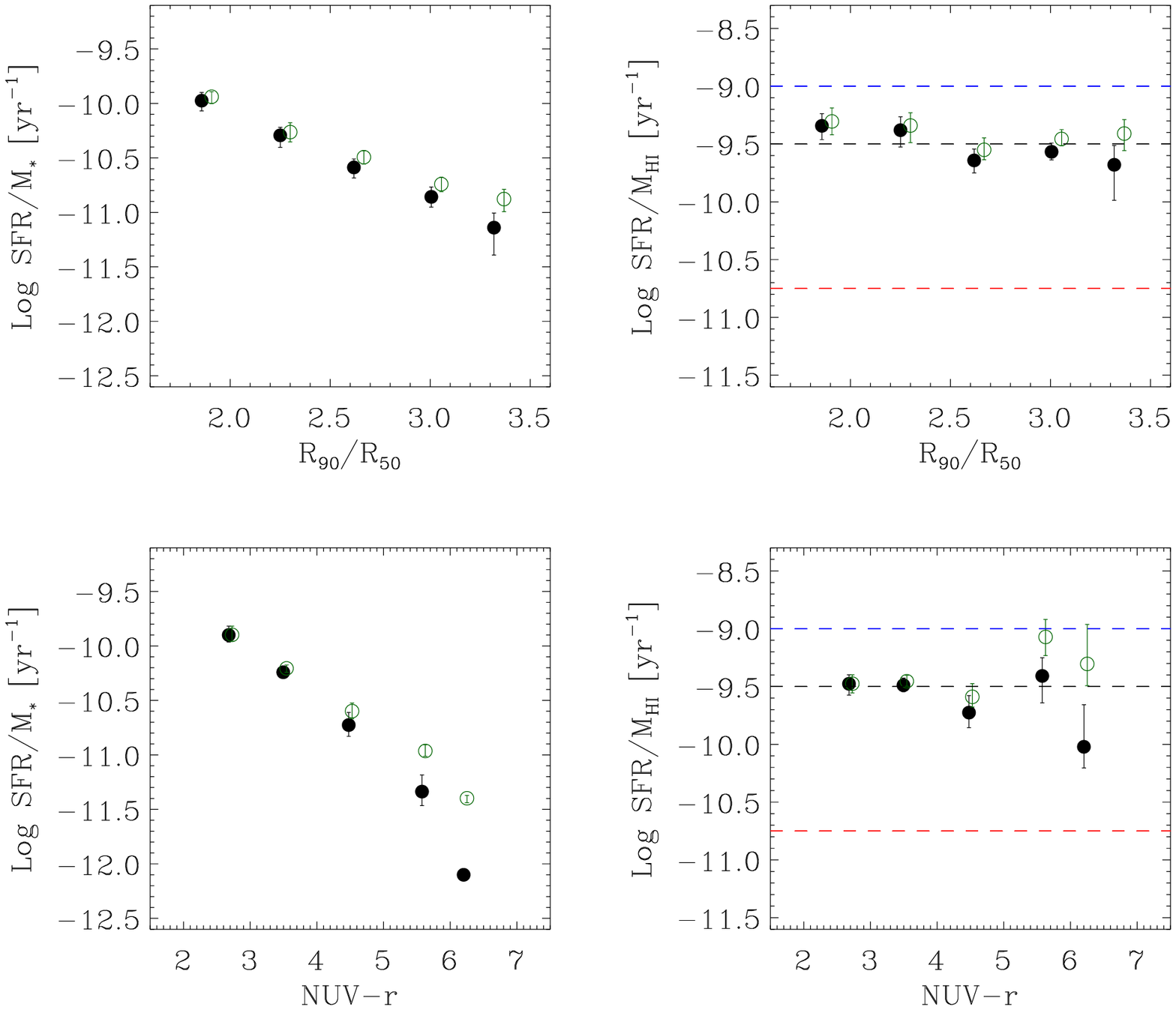}
\caption{Mean specific star formation rate (left) and star formation efficiency (right) as a function of observed quantities concentration, $R_{90}/R_{50}$ (top) and $NUV-r$, uncorrected for dust attenuation (bottom).   Symbols and colors as in Figure \ref{Fig:sfe_all}. \hi\ non-detections are given zero gas mass.  See text for details.} 
\label{Fig:sfe_all_obs}
\end{figure*}

To explore this further, in Fig \ref{Fig:sfe_all} and \ref{Fig:sfe_all_obs} we compare the global scaling relations of \ssfr\ and SFE and find that they each paint a very different picture.  In the left panel of each pair of figures we plot the binned specific star formation rate ($\Sigma SFR/\Sigma M_\star$) as a function of stellar mass, mass surface density, concentration and color (\Mstar, $\mu_\star$, $R_{90}/R_{50}$, and NUV-r).  In all cases, \ssfr~is steadily declining, typically by a factor of 10-30 across the range of the GASS sample.  In the right panels we show the average SFE ($\Sigma SFR/\Sigma M_{HI}$, where \hi\ non-detections are given zero \hi\ mass).   The average SFE is nearly flat, straddling $\sim 3 \times 10^{-9}$ yr$^{-1}$, corresponding to a gas depletion timescale of 3 Gyr.  Although in certain cases a small upwards or downward trend is suggested, these deviations are not statistically significant.   Values for these quantities are given in Tables \ref{Tab:SSFR} and \ref{Tab:SFE}. We have chosen to plot bin-averaged quantities where the numerator and denominator are summed separately in order to simplify the treatment of non-detections.  We have checked that this trend is also apparent for mean or median $SFR/M_\star$ and SFE measured for individual galaxies, the distribution of which is discussed in the next section.

This result is remarkable for two different reasons.  The first is that the timescale for gas consumption is nearly constant while the galaxy's specific star formation rate or ``building timescale'' is strongly dependent on stellar mass and correlated quantities.    

The second remarkable aspect is that the global average SFE that we measure for the GASS sample is very close to that observed locally for molecular gas in disks \citep[e.g.][using the THINGS survey, Walter et al. 2008]{Leroy2008}.  This suggests that the entire \hi~reservoir is being converted into stars at the same rate as the molecular gas.  It would appear to argue against a ``bottleneck'' in the flow of gas onto galaxies occurring, on average, at the interface between the atomic and molecular phase.  Instead it would appear that the gas-limiting step occurs prior to the \hi~phase; that quenching of the detectable cold gas is not responsible for regulating the star formation history of galaxies. We consider this result, and its relation to previous work, in the next section.

\section[]{Discussion}

\subsection{A nearly constant SFE}

We first start out by asking whether a constant volume-averaged SFE is surprising, particularly for galaxies in the GASS mass range?  While the global Schmidt law suggests a star formation efficiency that rises with gas surface density or varies inversely with dynamical or free-fall timescale \citep{Kennicutt1998}, more recent compilations of star formation laws \citep[e.g.,][]{Leroy2008} consider fixed giant molecular cloud star-forming efficiencies (with a varying atomic-to-molecular ratio) and  pressure regulation of the SFE and/or atomic-molecular ratio.  Because we measure neither the size of the gaseous and star-forming disk, nor the molecular phase, it is not easy to connect our global results to these theoretical predictions.  Global galaxy-averaged quantities sample a range of gas surface densities, timescales and conditions within the ISM.  Additionally, our mean SFEs combine measurements from galaxies with  range of morphological types, environments and presumably dark halo masses and spin parameters.  It is a considerable theoretical challenge to interpret this result on its own.

Observationally, low efficiencies have been measured for low-mass galaxies \citep{Geha2006}, LSB galaxies \citep{Boissier2008, Wyder2009}, and DLAs \citep{Wolfe2006}.  Conversely high efficiencies have been measured in starburst galaxies \citep[e.g.][]{Lehnert1996,Kennicutt1998} and in galaxies undergoing interactions \citep{Young1986, Solomon1988} and/or some form of environmental disturbance (e.g. stripping) \citep{Rose2009,KK04}. However, measurements of normal star-forming galaxies \citep[e.g.][]{Kennicutt1998} show that galaxies follow the global Schmidt law, implying a slowly varying SFE over the range of gas surface densities typically probed.

\citet{Young1986} and \citet{Devereux1991} noted a flat SFE across the Hubble sequence using $L(FIR)/M(H_2)$ as a tracer.    A similar result has been obtained by \citet{Boselli2001} and \citet{Boissier2001} who obtain a nearly constant SFE in a sample of normal spirals. 
More recently, \citet{Bothwell2009} suggested that the SFE is slowly increasing with galaxy luminosity (implying that more luminous galaxies have shorter gas consumption timescales), but those data, which only include \hi-detected galaxies, are nearly consistent with the constant values we have derived for the GASS sample.

\subsection{SFE vs. \ssfr}

We return to the comparative question:  Why does the \ssfr~drop sharply with \Mstar~and $\mu_\star$~while the SFE remains constant?  Here we consider two different scenarios and leave a more detailed analysis for future work.

\smallskip
\noindent{\it Internal regulation at the atomic-to-molecular transition:}  One possibility is that star formation is inhibited within the gas reservoir, at the sink point rather than the supply location.  Under the assumption that the efficiency of conversion of molecular gas into stars is nearly constant \citep[e.g.][]{Bigiel2008,Leroy2008}, regulation would then occur at the interface between the atomic and molecular phase.  Processes that can stabilize a gaseous disk, such as those proposed by \citet{Martig2009} are possible examples. As discussed in the introduction, although such a process can explain a decreasing \ssfr\ it may also result in a large reservoir of cold gas, leading to a decreasing SFE for galaxies that are actively being quenched.   Therefore it appears hard to reconcile internal regulation with the constant SFE and decreasing \ssfr\ vs. \Mstar\ that we observe (instead one might expect some correlation between SFE and \ssfr).  An alternative mechanism described in \citet{Hopkins2008}, with in-situ formation of molecular clouds in supershells, would also appear to be in conflict with our finding.

\smallskip
\noindent{\it Quenching or throttling of the \hi\ supply:}  A separate class of distinct processes are those that control the supply before (or quasi-instantaneously as) gas settles into the \hi\ phase.  This includes both the ``ejective feedback'' and ``preventive feedback'' quenching mechanisms discussed in \citet{Keres2009}.  For example, star formation (or AGN) may drive outflows, effectively ejecting gas from the system.  AGN feedback may heat infalling gas, or prevent its cooling, and allow the build up of a reservoir of gas in a bound hidden phase (e.g. ionized, warm-hot).   Although such a hidden phase has been invoked to balance accretion and star formation rates in the Milky Way \citep{Shull2009}, more generally quenched gas in this phase may or may not provide an additional supply for star formation through halo accretion, in that it may be permanently quenched, or may subsequently cool on a longer timescale \citep[e.g.,][]{Dekel2006}.   Additionally there are throttling scenarios that invoke varied gas accretion histories, essentially regulating the rate of infall onto the halo, with possible links to environmental conditions that lead to ``starvation'' or ``strangulation'' \citep{Larson1980}.   Finally, this category also includes delayed or staged accretion histories \citep{Boissier2001,Noeske2007}  where the effective accretion timescale varies with stellar mass.   In principle, what all of these have in common is that they can produce significant variation in \ssfr\ vs. stellar mass  while allowing the SFE to remain constant.

Interestingly, if interpreted in light of recent models explaining the star formation law in atomic and molecular gas \citep{Krumholz2008}, our result suggests that on average, most of the \hi\ in GASS galaxies must reside at surface densities at or near the atomic-to-molecular transition, with little or no systematic trend across the sample.  Our conclusion, that quenching/throttling mechanisms appear to better explain our data, has already been hinted at in previous work, most notably \citet{Larson1980} and \citet{Boissier2001} both of which identified similar observational trends in disk galaxy samples.  We note here that our average result does not preclude internal regulation from  taking place in some, but not all, of the massive galaxies in our sample.  It is possible that a ``conspiracy'' resulting from galaxies with low, internally quenched SFE being balanced by galaxies with high SFEs could reproduce the mean scaling relation.  We can investigate this by studying the actual distribution of SFE vs. stellar mass and other properties across the sample, which we do below.

\subsection{Distribution of SFE}

Scatter around the mean relations discussed above may provide insight into the episodic nature of gas accretion and quenching in GASS galaxies and we explore here the detailed distribution by plotting values and upper limits for individual galaxies in the DR1 sample.  We note here that while large uncertainties in some of the derived SFRs and SFEs can obscure visible dependencies, such upper limits and sources of error for low-SFR or low gas mass galaxies are unlikely to affect the global trends discussed in previous sections.  In general, their contribution to summed values is small.  On the other hand, random and systematic errors will also shift high SFR galaxies in these plots.  As discussed above, we've verified that our global results are robust with respect to such possible sources of error, but errors in SFRs, typically $\sim$0.3 dex, may produce some of the outliers on the plots described here.   Results quoted in this section should be treated as suggestive, motivating further study and improvements in the accuracy of these measures.

\begin{figure*}
\includegraphics[width=168mm]{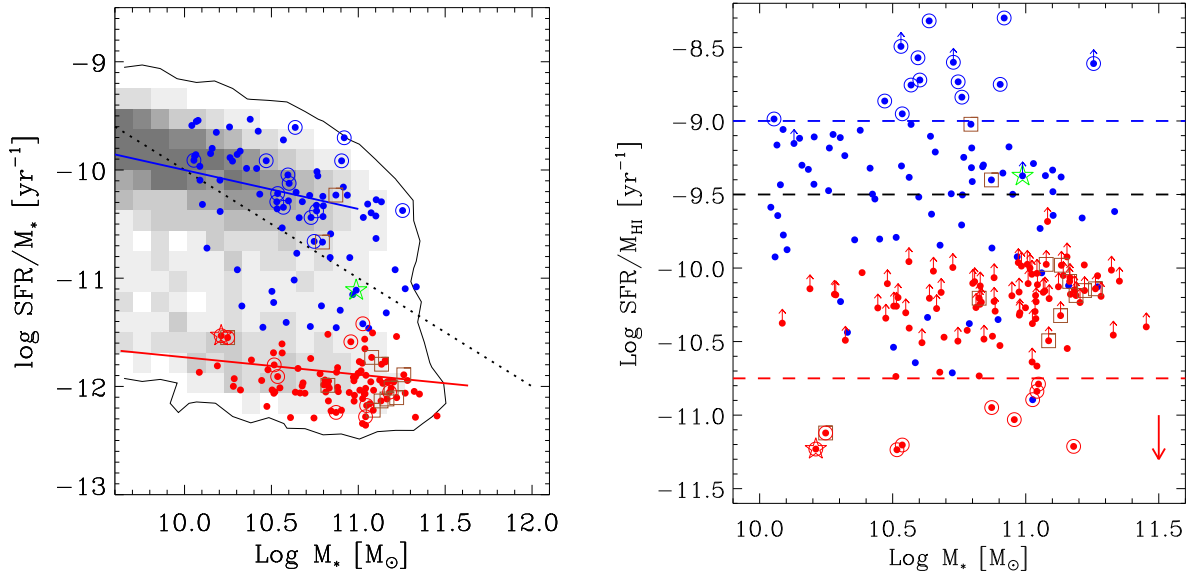}
\caption{Specific star formation rate  as a function of stellar mass (left) and star formation rate efficiency  as a function of stellar mass (right). Greyscale shows distribution of \ssfr~for galaxies in the local universe derived using volume-corrected UV-optical data from GALEX and SDSS \citep{Schiminovich2007}.  Blue/red points represent high/low specific star formation rate, split at $\log$ SFR/M$_\star$ = -11.5.  Large blue circles are high SFE galaxies (SFE $> 10^{-9.0}$ yr$^{-1}$).  Arrows indicate galaxies with no detection in \hi.  Large red circles are low SFE galaxies (SFE $< 10^{-10.75}$ yr$^{-1}$).  Green star is gas deficient object GASS 7050 and red star is gas-rich red galaxy GASS 3505 (see Paper I).  Brown squares indicate objects with nearby companions that may also be detected within the Arecibo beam.  In left panel the blue and red solid lines correspond to the star-forming sequence and the non-star forming locus of red galaxies. A constant star formation rate of 1 $M_\odot ~yr^{-1}$ is shown by the black dotted line.  In the right panel, the blue, black and red lines are used to show approximate locations of high, average and low SFE. Red arrow in the bottom right corner denotes amount SFR values of red points would move if 50\% of UV light came from evolved stars.}
\label{Fig:ssfr_sfe_mass}
\end{figure*}

In Figure \ref{Fig:ssfr_sfe_mass} we plot the specific star formation rate and star formation efficiencies as a function of stellar mass.  Also shown is the distribution of \ssfr~for galaxies in the local universe derived using volume-corrected UV-optical data from GALEX and SDSS \citep{Schiminovich2007}.   As expected, in any given stellar mass bin, GASS shows a similar distribution in \ssfr~as the volume-corrected GALEX+SDSS sample, consistent with the fact that GASS galaxies are selected purely based on stellar mass, with no other selection bias.  Our sample of massive galaxies is not purely passive; a significant fraction of GASS galaxies are on or close to the star-forming sequence.  In fact, many of the galaxies are forming stars at rates higher than 1 M$_\odot$ yr$^{-1}$.  Among the population of massive galaxies are some objects that have some of the highest SFRs in the local universe.  

There is a locus of galaxies near \ssfr $\sim 10^{-12}$ yr$^{-1}$, on the non-SF (red or dead) sequence.  As discussed above and in \citet{Schiminovich2007}, the \ssfr~for many of these galaxies is likely to be an overestimate due to the fact that UV light from evolved populations has not been subtracted when calculating star formation rates.  Although we do not label the red points as such, it is best to consider SFRs for these galaxies as an upper limit.

The right-hand side of Figure \ref{Fig:ssfr_sfe_mass}~shows the distribution of star formation efficiencies across the sample.  There is a broad spread of efficiencies in the sample with  implied gas consumption timescales ranging from less than 1 Gyr to over 100 Gyr.  In fact, this spread is indeed much larger than that typically seen for molecular gas within disk galaxies \citep[e.g. Figure 15 of][]{Leroy2008}.  We highlight in this and future figures those galaxies with exceptionally high and low star formation efficiencies.  Galaxies with SFE $> 10^{-9}$ yr$^{-1}$ are forming stars at rates that are high compared to their present atomic gas mass \citep[though still longer than likely dynamical timescales, consistent with ][]{Lehnert1996}.  Galaxies with SFE $<  10^{-10.75}$ yr$^{-1}$ appear to be passive in comparison to their current gas content.  GASS \hi~non-detections are shown with upward arrows, suggesting that their efficiencies could be higher.   The vast majority of the non-detections also have very low specific star formation rates which could suggest that estimates of their SFRs and SFEs are too high.  As a result, the red points that are \hi~non-detections may have higher {\it or} lower SFEs, and are poorly constrained in this diagram.  Overall, the broad scatter in star formation efficiencies vs. stellar mass does not point to a strong correlation between the two.

\begin{figure*}
\includegraphics[width=168mm]{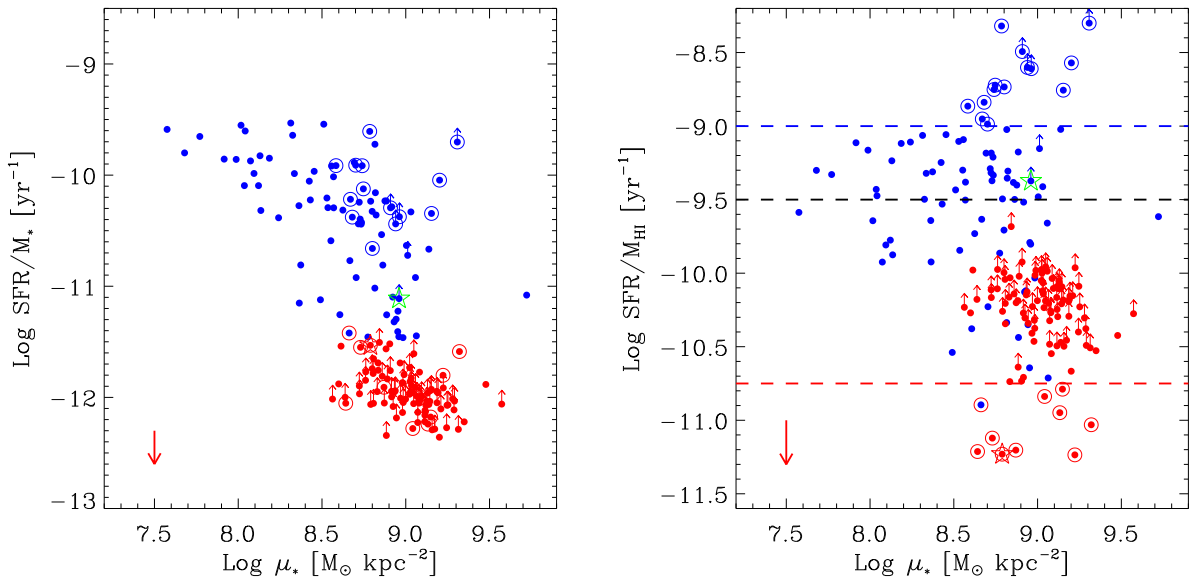}
\caption{Specific star formation rate  (left) and star formation rate efficiency  (right) as a function of stellar mass surface density.  Colors and symbols as in Figure \ref{Fig:ssfr_sfe_mass}.}
\label{Fig:logmu}
\end{figure*}

The high-SFE galaxies, indicated with large blue circles, are the galaxies with the shortest gas consumption timescales.  These galaxies are predominantly above the transition mass, $M_\star \sim 3\times10^{10}$ \msun.  Although these galaxies have elevated SFEs, they do not occupy an unusual location in the \ssfr~vs. $M_\star$ diagram, with most situated on or near the star-forming sequence.  This suggests that the reason why their star formation efficiency is high is because they have low gas mass for their given present averaged star-formation rate.

\begin{figure*}
\includegraphics[width=168mm]{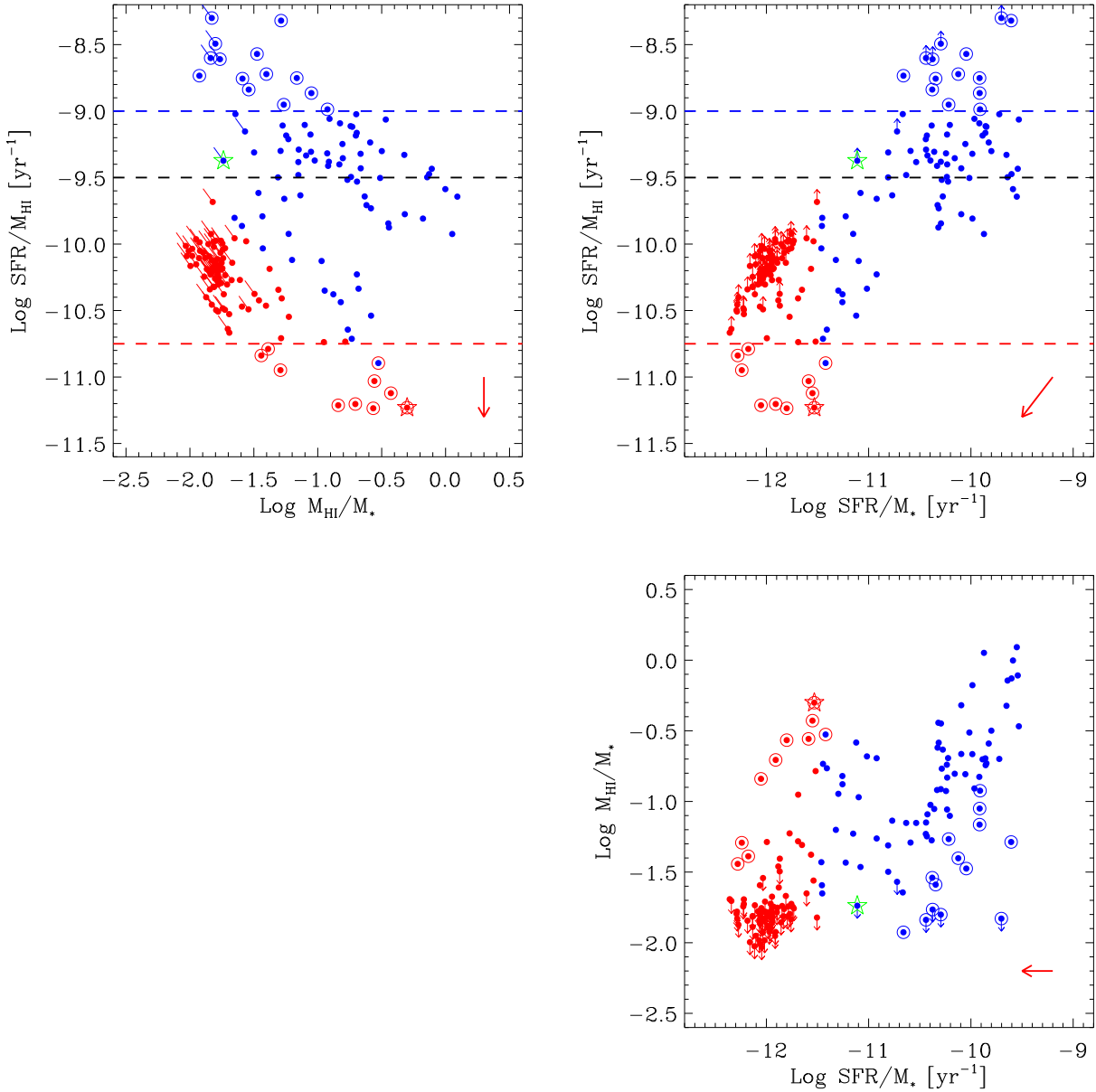}
\caption{Projected views of the 3-space of star formation efficiency vs. gas fraction vs. specific star formation rate.  SFE vs. \gfrac (top left) and  vs. \ssfr~(top right).  Lower panel shows \gfrac~ as a function of \ssfr. Colors and symbols as in Figure \ref{Fig:ssfr_sfe_mass}.  In the top left panel \hi\ masses are used to calculate both axes---upper limits are indicated by the short diagonal lines.}
\label{Fig:ssfr_gfrac_sfe}
\end{figure*}

Stellar mass surface density is known to display a strong correlation with specific star formation rate \citep{Kauffmann2003} and we show that the GASS sample displays a similar correlation in Figure \ref{Fig:logmu}.  Above a transition stellar mass surface density of $\mu_\star > 10^{8.5}$ \msun~kpc$^{-2}$ a much greater spread of \ssfr~is observed at a given $\mu_\star$.   This transition threshold has already been noted in Paper I and in previous work \citep{Kauffmann2003, Brinchmann2004}. On the other hand, the SFE, unlike the trend vs. stellar mass, shows a rather curious ``T-shaped'' distribution.  Below the $\mu_\star$~transition, the SFEs are nearly constant, straddling the line around SFE $=10^{-9.5}$ yr$^{-1}$.  Above the transition, the distribution in SFE becomes extremely broad, with very high and very low efficiencies present.  The highest SFE galaxies occupy the 'knee' of the \ssfr~vs. $\mu_\star$ diagram and might themselves represent a transition population \citep[see e.g.][]{Salim2007}.  The low SFE galaxies, on the other hand, appear to reside with passive galaxies. These galaxies host a sufficiently large gaseous reservoir that subsequent future star formation may lead to evolution off of the red sequence.  Our results highlight the fact that the combination of gas, star formation and stellar mass measurements can be useful for obtaining information about the possible future evolution of massive galaxies. We explore this further in Figure \ref{Fig:ssfr_gfrac_sfe} by investigating three plots that link gas fraction to \ssfr\ and SFE.  

In the first plot comparing SFE vs. \hi\ gas mass fraction (upper left), we see that high SFE galaxies have low gas fractions, with the converse being true for low SFE galaxies. Star-forming and green valley galaxies display a range of gas fractions from 2\% to 100\%.  We plot SFE vs. \ssfr\ (upper right), and find that the highest SFE galaxies do not have the highest specific star formation rates, but instead peak at slightly lower values. This suggests that a high SFE is not driven by a high SFR.     Not surprisingly, the gas fraction vs. \ssfr\ relation (lower right) shows that star-forming and green valley galaxies reveal correlated gas fraction and specific star formation rates with the extreme SFE galaxies lying off of this relation.    All of these plots, taken together, suggest that it is low \hi~content, as opposed to an excess in star formation rate, that is responsible for the majority of high SFE galaxies.      

\begin{figure*}
\includegraphics[width=135mm]{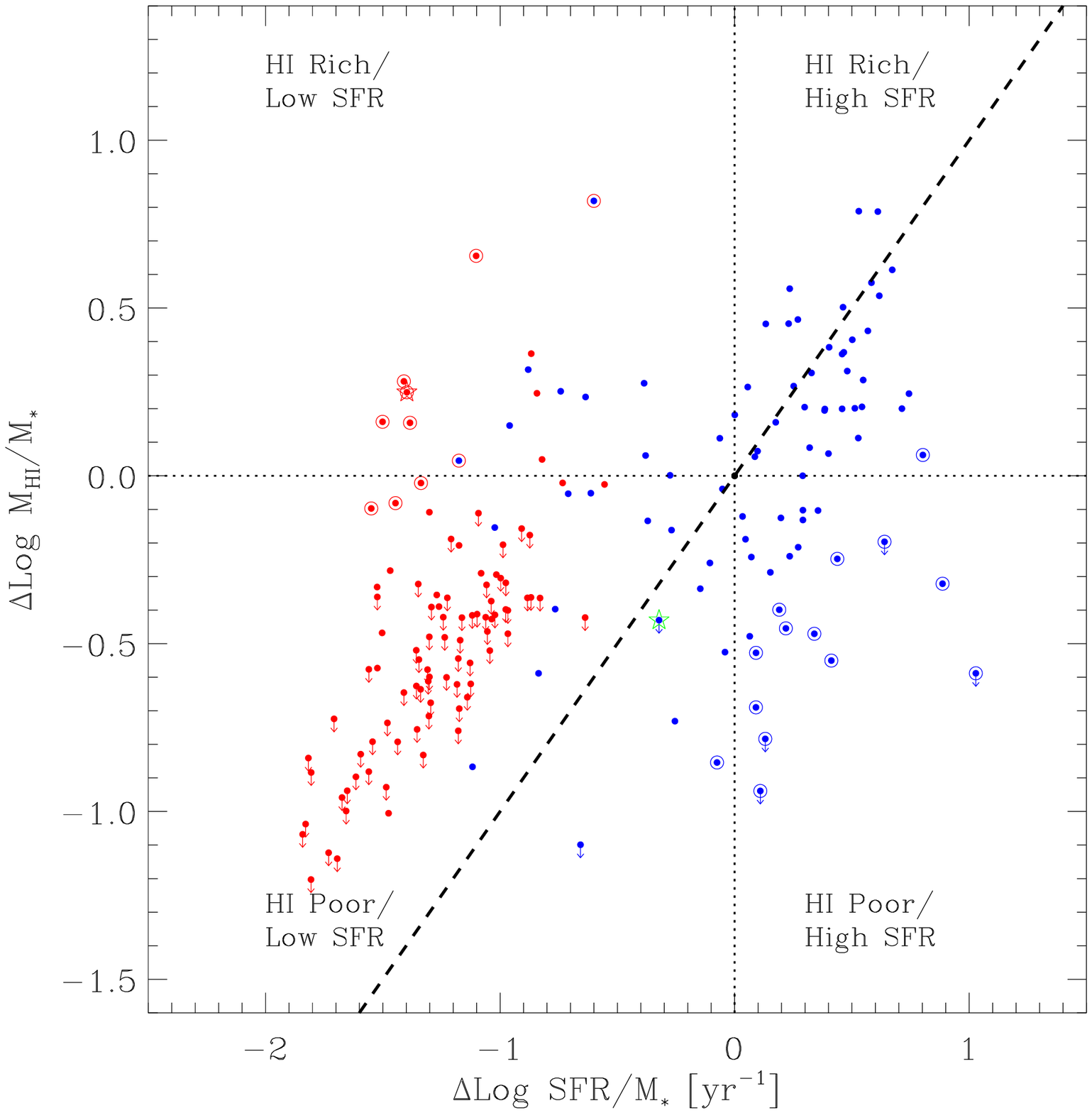}
\caption{Gas fraction excess vs. specific star formation rate excess.  Colors and symbols as in Figure \ref{Fig:ssfr_sfe_mass}. Zero-delta lines are plotted, as well as the line indicating $\Delta$ Log SFR/M$_\star$ = $\Delta$ Log M$_{\hi}$/M$_\star$. }
\label{Fig:delta_ssfr_gfrac}
\end{figure*}

Figures \ref{Fig:ssfr_sfe_mass}-\ref{Fig:ssfr_gfrac_sfe} can be used to isolate galaxies that show gas excess or deficiency when compared to their current star formation rates, useful for determining what causes high or low SFE in galaxies.    Figure \ref{Fig:delta_ssfr_gfrac} attempts to combine this diagnostic capability in one diagram, by plotting the offset of a galaxy's star formation rate and \hi~mass relative to the average value for a given stellar mass.  

We use data from this paper and Paper I and perform linear regression fits to the mean SFR and M$_{\hi}$ vs \Mstar.\footnote{We note that we have used the average SFR vs. \Mstar\ as opposed to the relationship between SFR and \Mstar\ along the star-forming sequence described in \citet{Schiminovich2007}.}  For the \hi\ masses we use the mean values for which the masses of non-detections have been set to the upper limit (Paper I,  Table 4, column 1). We find that
$$ \log <M_{\hi} (M_\star)> = 0.02 \log M_\star + 9.52 $$
and
$$ \log <SFR (M_\star) > = 0.15 \log M_\star - 1.5 $$
and define
$$\Delta \log \frac{M_{\hi}}{M_\star} = (\log M_{\hi} - \log <M_{\hi} (M_\star)> )   $$
$$\Delta \log \frac{SFR} {M_\star} = (\log SFR - \log <SFR (M_\star)> )  $$

In this delta gas-fraction vs. delta specific star formation rate plot four quadrants are delineated.  The top right quadrant contains those galaxies with high \ssfr~and \gfrac, in other words galaxies that  are gas-rich and actively star-forming. The majority of the GASS detections fall here along a line where gas fraction excess equals the \ssfr~excess which corresponds approximately to a line of constant SFE.    This quadrant is also likely to include gas-rich mergers, starbursts and high surface brightness galaxies with extended UV-disks \citep{Thilker2007}.  The upper left quadrant contains gas rich galaxies with lower than average star formation rates, including the low SFE galaxies highlighted above.  Low surface brightness galaxies and galaxies that have recently accreted gas might be found here.  The majority of galaxies identified as low SFE are forming stars below average rates, consistent with their being passive galaxies with large reservoirs of gas.  These galaxies do not appear to have extremely high gas masses and average star formation rates (which may be more typical of galaxies with lower stellar mass).  GASS3505, an unusually gas-rich galaxy discussed in Paper I, belongs to this class of object and is indicated with a red star in the figure.

The lower left quadrant contains gas-poor galaxies with below average SFR.  Passive galaxies are found in this quadrant.  Finally the lower right quadrant contains high SFE galaxies that are relatively gas poor but with above average SFR.  On this plot it becomes apparent that these galaxies are distributed across the quadrant and appear to have either high star formation rates when compared to their (typical) gas mass, or they have lower than average gas mass but typical star formation rates.   The most gas deficient in this category have possibly very recently experienced a process that disrupted gas flow and/or removed gas, such as starvation or stripping.  The latter scenario is likely to produce gas deficiencies even lower than we have observed.  Later stages of such galaxies might also be found in the third quadrant below or near the line of constant SFE. GASS 7050, identified in Paper I as an unusual gas-poor galaxy, appears to belong to this category and is indicated with a green star in the figure.   

Lastly, we return to the question of whether or not the scatter around our mean scaling relations suggests a diversity in the processes that trigger and quench star formation in the GASS sample.   Without question the GASS sample does not show a tight distribution in SFEs, and it is tempting to reconsider internal quenching mechanisms for at least some fraction of systems.    Alternately, our results may rule out scenarios where the inflow of material is occurring as a steady flow (or drizzle) onto the galaxy.  Instead, the scatter may be an indication that accretion of \hi\ is episodic, with infalling gas arriving in larger discrete chunks.   It may suggest that large scale processes play a role in regulating the growth and evolution of gas and star formation in galaxies. 

Future work is being planned to investigate the large- and small-scale environment of GASS galaxies, and in particular the outliers, which should reveal whether enviromental processes are driving this evolution.  Some galaxies with high star formation rates may be undergoing a merger or interaction that is driving up the star formation rate in these galaxies and it will also be interesting to investigate the connection between signs of interaction, merging or other disturbance and location on this diagram.

\section[]{Conclusions}

We use measurements of the \hi\ content, stellar mass and star formation rates in $\sim 190$ massive galaxies with $M_\star  >10^{10}$ \msun, obtained from the Galex Arecibo SDSS survey described in Paper I \citep{Catinella2010} to explore the global scaling relations associated with the ratio \sfe~, which we call the \hi~based star formation efficiency (SFE).   We find that:

\renewcommand{\labelenumi}{\arabic{enumi}.}

\begin{enumerate}
\item We can measure the volume-averaged \hi\ mass density and SFR density for the GASS sample.  GASS galaxies account for $36\pm5$ percent of the total \hi\ mass density and $47\pm5$ percent of the SFR mass density.

\item Unlike the mean specific star formation rate, which decreases with stellar mass and stellar mass surface density, the star formation efficiency remains relatively constant across the sample with a value close to $SFE = 10^{-9.5}$ yr$^{-1}$ (or an equivalent gas consumption timescale of $\sim 3 \times 10^9$ yr).  We find little variation in SFE with stellar mass, stellar mass surface density, $NUV-r$ color and concentration.  We interpret these results as an indication that external processes or feedback mechanisms that control the gas supply are important for regulating star formation in massive galaxies.

\item Approximately 5\% of the sample shows high efficiencies with SFE $> 10^{-9}$ yr$^{-1}$, and we suggest that this is very likely due to a deficiency of cold gas rather than an excess star formation rate.  Conversely, we also find a similar fraction of galaxies that appear to be gas-rich for their given specific star-formation rate, although such galaxies show both a higher than average gas fraction and lower than average specific star formation rate.  Both of these populations are plausible candidates for ``transition'' galaxies, with potential for a change (either decrease or increase) in their specific star formation rate in the near future.  

\end{enumerate}

\section*{Acknowledgments}

We thank the Arecibo staff, in particular Phil Perillat,
Ganesan Rajagopalan and the telescope operators for their assistance,
and Hector Hernandez for scheduling the observations.   
R.G. and M.P.H. acknowledge support from NSF grant AST-0607007 and
from the Brinson Foundation.

The Arecibo Observatory is part of the National Astronomy and
Ionosphere Center, which is operated by Cornell University under a
cooperative agreement with the National Science Foundation. 

GALEX (Galaxy Evolution Explorer) is a NASA Small Explorer, launched
in April 2003. We gratefully acknowledge NASA's support for
construction, operation, and science analysis for the GALEX mission,
developed in cooperation with the Centre National d'Etudes Spatiales
(CNES) of France and the Korean Ministry of Science and Technology.

Funding for the SDSS and SDSS-II has been provided by the Alfred
P. Sloan Foundation, the Participating Institutions, the National
Science Foundation, the U.S. Department of Energy, the National
Aeronautics and Space Administration, the Japanese Monbukagakusho, the
Max Planck Society, and the Higher Education Funding Council for
England. The SDSS Web Site is http://www.sdss.org/.

The SDSS is managed by the Astrophysical Research Consortium for the
Participating Institutions. The Participating Institutions are the
American Museum of Natural History, Astrophysical Institute Potsdam,
University of Basel, University of Cambridge, Case Western Reserve
University, University of Chicago, Drexel University, Fermilab, the
Institute for Advanced Study, the Japan Participation Group, Johns
Hopkins University, the Joint Institute for Nuclear Astrophysics, the
Kavli Institute for Particle Astrophysics and Cosmology, the Korean
Scientist Group, the Chinese Academy of Sciences (LAMOST), Los Alamos
National Laboratory, the Max-Planck-Institute for Astronomy (MPIA),
the Max-Planck-Institute for Astrophysics (MPA), New Mexico State
University, Ohio State University, University of Pittsburgh,
University of Portsmouth, Princeton University, the United States
Naval Observatory, and the University of Washington.

\appendix

\vfill
\clearpage

\begin{table}
 \caption{Mean specific star formation rates within the GASS sample}
 \label{Tab:SSFR}
 \begin{tabular}{@{}lcccccc}
\hline
\hline
 x & $<$x$>$ & Log $<$SFR/M$_\star$ yr $>^a$ &  Log $<$SFR/M$_\star yr>^b$ & \\
  \hline
Log M$_\star$ &  10.17 &  -10.11 $\pm$   0.06 & -10.08 $\pm$   0.05 \\
 &  10.49 &  -10.47 $\pm$   0.11 & -10.41 $\pm$   0.09 \\
 &  10.76 &  -10.62 $\pm$   0.10 & -10.55 $\pm$   0.08 \\
 &  11.05 &  -11.08 $\pm$   0.09 & -10.90 $\pm$   0.06 \\
 &  11.29 &  -11.17 $\pm$   0.17 & -11.01 $\pm$   0.07 \\
  \hline
Log $\mu_\star$ &   8.13 &   -9.92 $\pm$   0.16 &  -9.92 $\pm$   0.12 \\
 &   8.46 &  -10.33 $\pm$   0.11 & -10.25 $\pm$   0.08 \\
 &   8.77 &  -10.61 $\pm$   0.07 & -10.53 $\pm$   0.05 \\
 &   9.05 &  -11.26 $\pm$   0.14 & -10.98 $\pm$   0.07 \\
 &   9.27 &  -11.38 $\pm$   0.33 & -11.07 $\pm$   0.14 \\
  \hline
R$_{90}$/R$_{50}$ &   1.85 &  -10.07 $\pm$   0.12 & -10.01 $\pm$   0.09 \\
 &   2.28 &  -10.38 $\pm$   0.12 & -10.34 $\pm$   0.12 \\
 &   2.61 &  -10.63 $\pm$   0.08 & -10.52 $\pm$   0.06 \\
 &   3.02 &  -10.96 $\pm$   0.13 & -10.81 $\pm$   0.09 \\
 &   3.32 &  -11.45 $\pm$   0.22 & -11.02 $\pm$   0.07 \\
  \hline
NUV$-$r &   2.71 &   -9.93 $\pm$   0.09 &  -9.92 $\pm$   0.09 \\
 &   3.54 &  -10.33 $\pm$   0.06 & -10.29 $\pm$   0.05 \\
 &   4.50 &  -10.83 $\pm$   0.10 & -10.67 $\pm$   0.06 \\
 &   5.60 &  -11.38 $\pm$   0.19 & -11.00 $\pm$   0.07 \\
 &   6.18 &  -12.09 $\pm$   0.02 & -11.37 $\pm$   0.03 \\
  \hline
  \hline
 \end{tabular}

\noindent ~$^a$ SFR is NUV-based with dust-correction applied to galaxies with D$_n$(4000) $<$ 1.7 (SFR$_{NUV,DC,D4NCUT}$). Uncertainties do not include possible systematic errors in SFR.

\noindent ~$^b$ SFR is NUV-based with dust-correction applied to all galaxies (SFR$_{NUV,DC}$).  Uncertainties do not include possible systematic errors in SFR.

 \medskip
\end{table}

\begin{table}
 \caption{Mean star formation efficiencies of the GASS sample}
 \label{Tab:SFE}
 \begin{tabular}{@{}lcccccc}
\hline
\hline
 x & $<$x$>$ & Log $<$SFR/M$_{HI}>^a$ & Log $<$SFR/M$_{HI}>^b$ & \\
  \hline
Log M$_\star$ &  10.17 &   -9.58 $\pm$  -0.05 &  -9.54 $\pm$  -0.04 \\
 &  10.49 &   -9.52 $\pm$  -0.10 &  -9.45 $\pm$  -0.09 \\
 &  10.76 &   -9.49 $\pm$  -0.11 &  -9.41 $\pm$  -0.11 \\
 &  11.05 &   -9.74 $\pm$  -0.10 &  -9.55 $\pm$  -0.08 \\
 &  11.29 &   -9.60 $\pm$  -0.22 &  -9.46 $\pm$  -0.13 \\
  \hline
Log $\mu_\star$ &   8.13 &   -9.64 $\pm$  -0.13 &  -9.61 $\pm$  -0.11 \\
 &   8.46 &   -9.60 $\pm$  -0.12 &  -9.54 $\pm$  -0.11 \\
 &   8.77 &   -9.63 $\pm$  -0.07 &  -9.55 $\pm$  -0.05 \\
 &   9.05 &   -9.66 $\pm$  -0.15 &  -9.38 $\pm$  -0.11 \\
 &   9.27 &   -9.50 $\pm$  -0.43 &  -9.18 $\pm$  -0.35 \\
  \hline
R$_{90}$/R$_{50}$ &   1.85 &   -9.48 $\pm$   0.07 &  -9.43 $\pm$   0.08 \\
 &   2.28 &   -9.39 $\pm$   0.17 &  -9.35 $\pm$   0.15 \\
 &   2.61 &   -9.68 $\pm$   0.12 &  -9.58 $\pm$   0.09 \\
 &   3.02 &   -9.64 $\pm$   0.10 &  -9.49 $\pm$   0.08 \\
 &   3.32 &   -9.94 $\pm$   0.24 &  -9.48 $\pm$   0.17 \\
  \hline
NUV$-$r &   2.71 &   -9.51 $\pm$   0.10 &  -9.51 $\pm$   0.10 \\
 &   3.54 &   -9.52 $\pm$   0.05 &  -9.46 $\pm$   0.05 \\
 &   4.50 &   -9.83 $\pm$   0.13 &  -9.66 $\pm$   0.09 \\
 &   5.60 &   -9.48 $\pm$   0.22 &  -9.09 $\pm$   0.12 \\
 &   6.18 &   -9.95 $\pm$   0.16 &  -9.21 $\pm$   0.17 \\
  \hline
  \hline
 \end{tabular}
 
{  \noindent ~$^{a, b}$ SFR and uncertainties} as in Table \ref{Tab:SSFR}.
 \medskip
\end{table}

\label{lastpage}

\end{document}